# Ramsey-comb spectroscopy with intense ultrashort laser pulses

Jonas Morgenweg, Itan Barmes, Kjeld S. E. Eikema*

LaserLaB Amsterdam, VU University, De Boelelaan 1081, 1081HV Amsterdam, The Netherlands.
*Correspondence to: K.S.E.Eikema@vu.nl

**Optical frequency combs based on mode-locked lasers have revolutionised the field of metrology and precision spectroscopy by providing precisely calibrated optical frequencies and coherent pulse trains [1, 2]. Amplification of the pulsed output from these lasers is very desirable, as nonlinear processes can then be employed to cover a much wider range of transitions and wavelengths for ultra-high precision, direct frequency comb spectroscopy [3, 4]. Therefore full repetition rate laser amplifiers [5, 6] and enhancement resonators [7, 8] have been employed to produce up to microjoule-level pulse energies [9]. Here we show that the full frequency comb accuracy and resolution can be obtained by using only two frequency comb pulses amplified to the millijoule pulse energy level, orders of magnitude more energetic than what has previously been possible. The novel properties of this approach, such as cancellation of optical light-shift effects, is demonstrated on weak two-photon transitions in atomic rubidium and caesium, thereby improving the frequency accuracy up to thirty times.**

As an alternative to full repetition rate amplification and cavity enhancement of frequency combs (FC), direct amplification of selected FC pulses allows for much higher pulse energies and wavelength tunability. By amplifying two FC pulses and subsequent harmonic upconversion, precision spectroscopy in the extreme ultra-violet near 51 nm has been demonstrated [10]. However, in [10] the FC resolution was sacrificed because only two *consecutive* FC pulses could be amplified, and phase shift effects during the amplification process compromised the FC accuracy. To realise both frequency comb resolution and accuracy in conjunction with mJ-pulse energies, we developed the method of Ramsey-comb spectroscopy. This method is based on a *series* of excitations with two selectively amplified frequency comb laser pulses, which can be varied in delay over a wide range without affecting the optical phase.

Traditionally, excitation of atoms or molecules with two short and phase-coherent laser pulses is known as Ramsey spectroscopy [11, 12]. The pulses induce two excitation contributions that interfere depending on the delay time ($\Delta t$) and a possible additional phase shift between the pulses ($\Delta \phi$, e.g. from a pulse-amplification process). For a two-level atom with transition frequency $f_i$, the excited state population will exhibit an oscillatory behaviour when $\Delta t$ is changed, proportional to $1+\cos(2\pi f_i \Delta t + \Delta \phi)$ (see Fig. 1a and Supplementary Fig. S1). If this signal is measured over a few oscillation periods as a function of $\Delta t$ (a Ramsey-scan), then the transition frequency can be determined very precisely, provided that $\Delta t$ and $\Delta \phi$ are known. A larger $\Delta t$ leads to a more accurate determination of the transition frequency $f_i$. However, Ramsey spectroscopy based on a single scan can only measure one isolated transition at a time, and is sensitive to errors in $\Delta \phi$ [10].

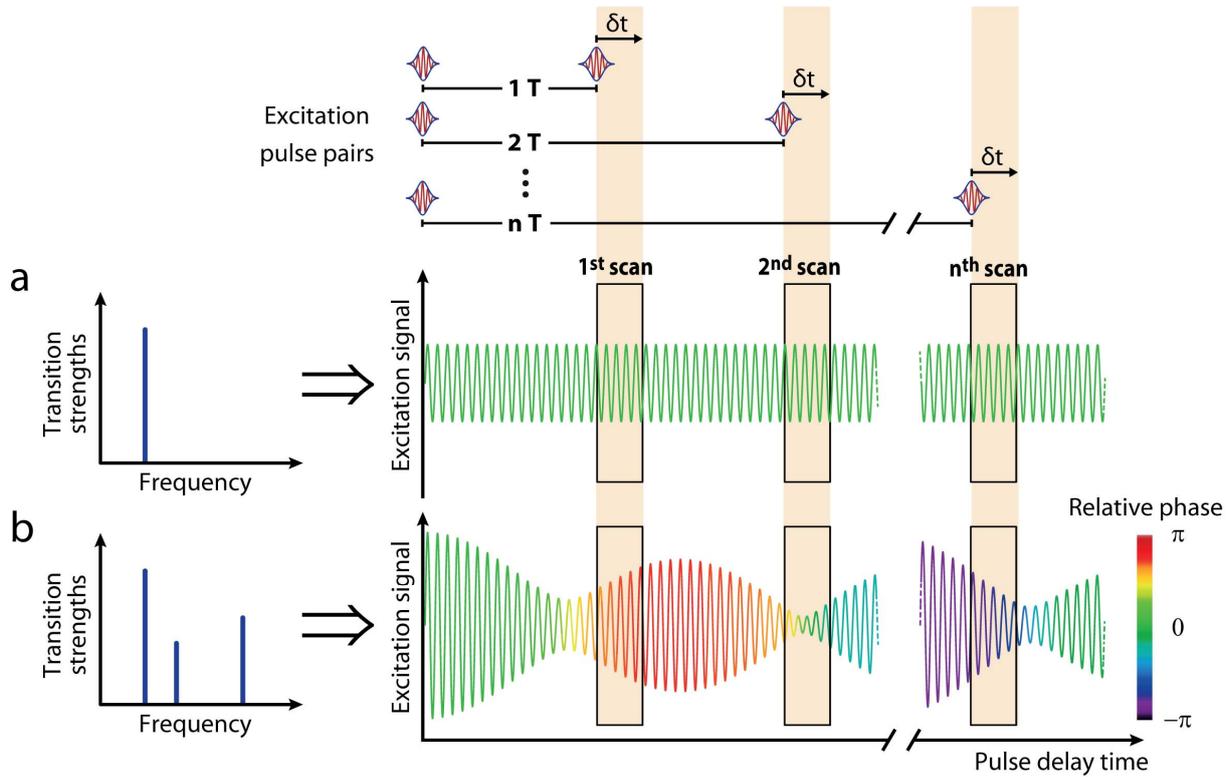

**Figure 1: The principle of Ramsey-comb spectroscopy.** An atomic system is excited with two coherent laser pulses at a widely tunable and accurate delay, provided by a frequency comb. The laser pulses sample the excited population signal by a short Ramsey-scan over $\delta t$ at macro-delays that are an integer ($n$) multiple of the comb repetition time $T$. From these scans the transition frequencies and strengths can be reconstructed with high precision. **a,** In the case of only one resonance, the excitation signal undergoes a single cosine modulation of constant amplitude known as Ramsey fringes. **b,** If multiple transitions are excited simultaneously, the resulting signal will exhibit a complex amplitude and phase pattern. The phase evolution is visualised in colour relative to the single transition in part **a**.

Instead, in Ramsey-comb spectroscopy a series of individual Ramsey-scans are performed using coherently-amplified pulse pairs derived from a FC laser. The coarse delay of the pulse pairs can be changed in steps of the FC repetition time $T$, while fine tuning for a Ramsey scan is achieved by small adjustments of $T$ itself. As a result, we obtain a "comb" of Ramsey signals, with three fundamental properties.

First, the FC provides a precisely calibrated absolute time axis and phase control over a wide range of pulse delays (> microseconds), enabling very precise frequency determination.

Second, if a constant phase shift $\Delta\phi$ affects the Ramsey signals, then it can be identified as a common effect in all the signals recorded at different time delays. It therefore drops out of the analysis and the full frequency comb accuracy is recovered. Note that this includes light-induced phase shifts due to AC-Stark and similar effects [13], which often lead to frequency errors in (frequency comb) spectroscopy.

Third, by probing the excited state population over longer periods, multiple transitions can be measured simultaneously by observing a beating between the individual cosine contributions from each resonance at frequency $f_i$ with transition strength $A_i$. The multi-transition signal will be proportional to:

$$S = \sum_i A_i \left[1 + \cos(2\pi f_i \Delta t + \Delta\phi)\right]. \tag{1}$$

As an example, the expected upper state population signal for three transitions as a function of the inter-pulse delay is schematically depicted in Fig. 1b. It can be seen that analogous to the superposition of sound waves from slightly detuned tuning forks, the excitation signal exhibits a characteristic beating pattern. The excitation oscillations are related to those observed in traditional Fourier-transform spectroscopy [14], or similar methods with pulsed lasers based on physical optical delay lines [15, 16]. However, in Ramsey-comb spectroscopy the FC source provides an absolute time axis for the pulse delay $\Delta t$, and this for time scales many orders of magnitude larger than any physical delay line can provide. Moreover, the individually acquired Ramsey-scans result in accurate information on the phase of the complex delay-dependent signal, as visualised by the colour-gradient of the signal trace in Fig. 1b. This phase information is robust against fluctuations of signal strength and encodes both the transition frequencies and strengths. The underlying resonances can therefore be obtained very accurately from a straightforward fit of the phase according to formula (1), without complications introduced by line shapes in the frequency domain (more details on the fitting procedure is found in the Supplementary Information).

The frequency domain spectrum can be calculated as well from the Ramsey scans by a discrete Fourier-transform over all measured delay zones. These spectra are subtly different from normal frequency comb spectroscopy, but enable straightforward identification of the transitions, and provide good starting values for the phase fit performed on Ramsey signals in the time domain (see Supplementary Information).

Experimentally, we obtain Ramsey-comb pulse pairs from a fully referenced Ti:sapphire FC laser, operating near 760 nm with a repetition rate of $f_{rep} \approx 128$ MHz. Two pulses from this comb laser are parametrically amplified more than a million times up to 5 mJ. The parametric amplifier supports broadband operation [17], but for this experiment only a 5 nm wide part of the spectrum is selected. The pulse delay of the amplified FC pulses is determined by the pump laser as visualised in Fig. 2. Only the FC pulses overlapping temporally with the high-energy pump pulses are amplified in the parametric amplifier. We verified that there is no delay-dependent phase shift introduced in the amplification process within an accuracy of <1/1000[th] of an optical cycle, based on spectral interferometry with the original frequency comb pulses [18].

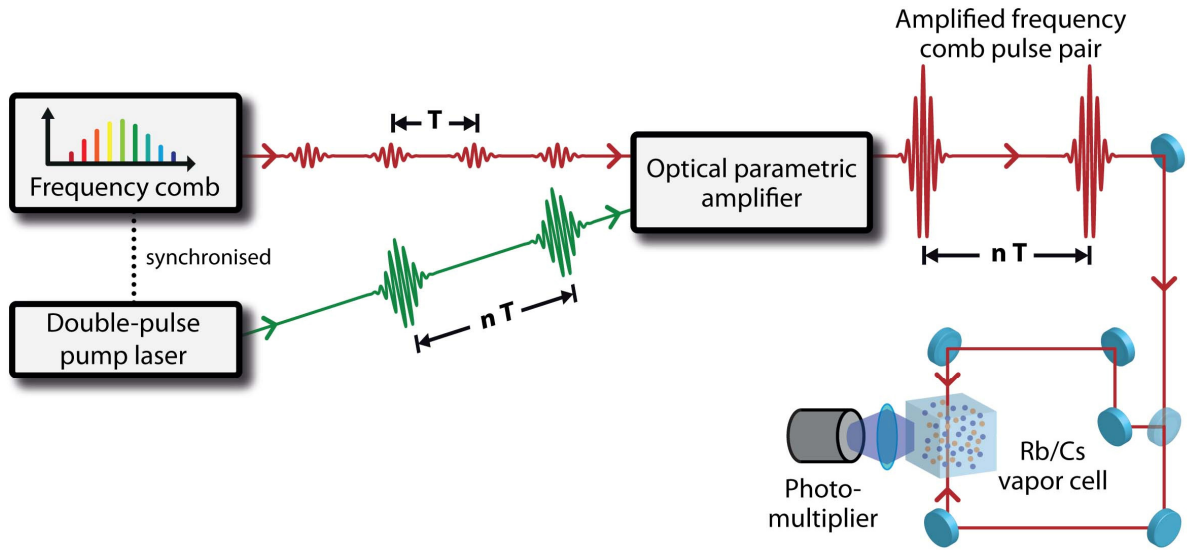

**Figure 2: Schematic of the experimental setup.** A high-energy pump pulse pair selectively amplifies two pulses from a frequency comb laser pulse train. The macro-delay between the pump pulses and hence the amplified frequency comb pulses can be changed in steps of the cavity round-trip time $T$ = 7.8 ns (where $n$ is an integer number). The amplified pulse pairs are then split into counter-propagating copies to perform Doppler-reduced two-photon spectroscopy in a cell containing a mixture of atomic rubidium and caesium vapour. Signal is detected by monitoring fluorescence decay of excited atoms with a photo-multiplier tube.

To demonstrate the capabilities of Ramsey-comb spectroscopy, the amplified FC pulse pairs are used to perform non-resonant two-photon spectroscopy in an atomic vapour cell (Fig. 2). Although the investigated transitions are very weak, no focusing of the laser beam (which has a diameter of 3 - 6 mm depending on experimental conditions) is required because of the high pulse energy. At every macro-delay step $n$, the inter-pulse delay is scanned in steps of a few hundred attoseconds by small changes of the repetition rate of the FC oscillator. This results in Ramsey scans consisting of a few oscillations of the fluorescence signal, which is recorded with a photo-multiplier. Further experimental details can be found in the Methods section.

A typical measurement for rubidium and caesium is shown in Fig. 3a; the signals are corrected for a constant background in the vertical direction. The change in Ramsey-signal amplitude between the macro-delay steps ($T$ = 7.8 ns) is a direct result of the beating of the individual fluorescence signals from simultaneously excited transitions. Because these contrast changes appear on a nanosecond time scale, there is only a negligible effect on the signal amplitude within one Ramsey-scan of ~3 fs length. For longer delays (higher $n$), there is an additional, general reduction in contrast due to the residual Doppler-effect and spontaneous decay of the excited states. In the case of e.g. rubidium this limits the useable delay to about 345 ns ($n = 44$) due to the upper state lifetime of 88 ns [19].

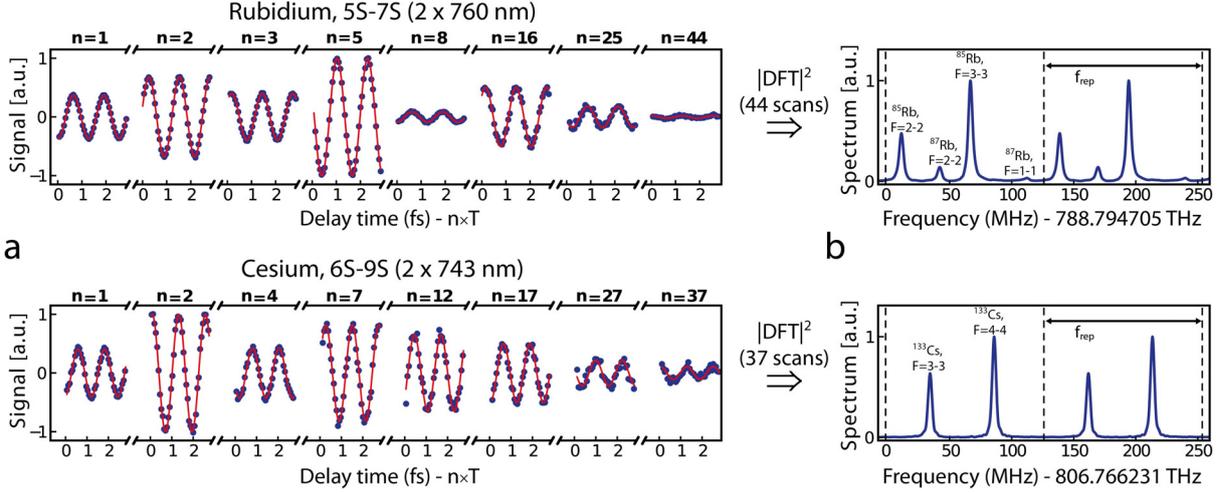

**Figure 3: Experimental demonstration of Ramsey-comb spectroscopy. a,** Upper part: selection of measured Ramsey-comb signal of two-photon 5S-7S transitions in atomic $^{85}$Rb and $^{87}$Rb, at macro-delays of $nT$ ($T = 7.837146$ ns). Lower part: similarly for the 6S-9S transition in $^{133}$Cs. For each delay step $n$, the inter-pulse delay $\Delta t$ was fine-adjusted over a range of $\delta t \approx 3$ fs to record a few oscillations of the signal beating pattern, such that $\Delta t = nT + \delta t$. The solid line represents a sinusoidal fit. **b,** Calculated spectrum based on the discrete Fourier-transform (DFT) of the time domain signal from a total of 44 (rubidium) and 37 (caesium) Ramsey-scans. The spectral patterns repeat with a period of ~128 MHz (=1/$T$) and is used only for identification of the transitions (see text).

Regarding the 5S-7S transition in $^{85}$Rb we arrive at the transition frequency before hyperfine splitting ("centre of gravity frequencies", $f_{cog}$) and hyperfine $A$ constants of $f_{cog}$ = 788,796,960,604(5) kHz and $A_{7S}$ = 94,684(2) kHz (based on 28 datasets). For the same transition in $^{87}$Rb we find $f_{cog}$ = 788,797,092,128(7) kHz and $A_{7S}$ = 319,761(6) kHz. The uncertainties are a combination of statistical and systematic errors (see Supplementary Information for more details). Because of small laser power drifts up to a few percent during the measurements, the AC-Stark (light) shift effect was not perfectly cancelled. However, still an effective ~50 times suppression was accomplished, leading to only small residual AC-Stark shift corrections of a few kHz.

The measurements presented here are in good agreement with previous experiments [20, 21], and also of the same accuracy as the best determination recently obtained with full repetition rate comb excitation, employing strong focusing of the nJ-level laser pulses and coherent control [21]. This confirms that Ramsey-comb spectroscopy can be at least as accurate as full repetition rate frequency comb spectroscopy, but at many orders of magnitude higher pulse energy.

The advantage of high pulse energies becomes apparent when Ramsey-comb spectroscopy is applied on much weaker transitions, such as the investigated 6S-9S transition in $^{133}$Cs. As shown in Fig. 3, a strong signal is obtained without any need for resonant enhancement by an intermediate level. From the analysis we find $f_{cog}$ = 806,761,363,429(7) kHz and $A_{9S}$ = 109,999(3) kHz, which is thirty times more accurate than previous direct frequency comb measurements on this transition [22]. The Ramsey-comb method therefore outperforms traditional forms of continuous wave or FC laser spectroscopy on transitions that are too weak to be easily excited with unamplified frequency comb pulses.

Based on parametric amplification, Ramsey-comb spectroscopy combines high frequency precision with wide wavelength coverage at mJ-level pulse energy. Because of the high peak energy, the frequency range of this method can straightforwardly and efficiently be extended via nonlinear crystals to the ultraviolet, or with high-harmonic generation in a gas jet to the extreme ultraviolet [23] (taking $T$ > 100 ns to avoid phase shifts from ionisation in the gas jet). Therefore there are many interesting targets for the Ramsey-comb method, such as the 1S-2S two-photon transition in He$^+$ to provide new information on the proton-size puzzle [24, 25], or the two-photon X-EF transition in molecular hydrogen to put tighter constraints on speculative 5$^{th}$ forces beyond the Standard Model [26].

**Methods:**
The FC laser providing the seed pulses for the parametric amplifier is a home-built, Kerr-lens mode-locked Ti:sapphire oscillator. Both its repetition rate and carrier-to-envelope phase are locked to an atomic Rb-clock disciplined by the Global Positioning system (fractional accuracy better than 2×10$^{-12}$ for averaging times larger than 100 s). The oscillator emits pulses of 6 nJ energy, at a repetition time of 7.8 ns, and with a spectral bandwidth of ~40 nm centred at 760 nm. Before amplification, the pulses are stretched to 10 ps, by the combined effect of clipping the spectrum to about 5 nm around the desired wavelength and the application of ~690,000 fs$^2$ of group delay dispersion. The stretched FC pulses are selectively amplified in an optical parametric amplifier to the mJ-level by a high-energy 532 nm pump-pulse pair. The pump pulses originate from a separate, passively mode-locked Nd:YVO$_4$ oscillator, which is electronically synchronised to the Ti:sapphire FC oscillator at the same $f_{rep}$ ≈ 128 MHz. Via programmable pulse-pickers two pulses are selected from the pump oscillator pulse train. These pulses are amplified to 40 mJ with an ultra-high gain Nd:YVO$_4$ pre-amplifier system [27, 28] and a Nd:YAG post amplifier, and subsequently frequency-doubled to 24 mJ at 532 nm. The parametric amplifier then produces amplified FC pulse pairs up to 5 mJ energy at a repetition frequency of 28 Hz, which therefore determines the repetition rate of the total experiment. During the amplification process, both pump pulses travel exactly the same optical path, assuring that their wavefronts are equal on a sub-milliradian level. This is essential because the parametric amplification is a highly-nonlinear process and the amplified signal phase is very sensitive to differences in wavefronts [18].

The Doppler-reduced two-photon spectroscopy is performed in a cell containing a mixture of rubidium and caesium vapour, heated to ~50°C. Background signal originating from single-sided excitation is strongly suppressed because of the chirp of the amplified FC pulses [29], combined with the use of quarter-wave plates to generate circular polarised light. The signal is proportional to the number of excited atoms as a function of inter-pulse delay, and is recorded by monitoring the fluorescence decay (420 - 459 nm) to the ground state after the second excitation pulse.

**Acknowledgments:**

We gratefully acknowledge financial support by the Netherlands Organization for Scientific Research (NWO) through VICI grant 680-47-310, the EC's Seventh Framework Programme LASERLAB EUROPE (JRA INREX), and the Foundation for Fundamental Research on Matter (FOM) for support through its program "Broken Mirrors and Drifting Constants". We also would like to thank J. Bouma and R. Kortekaas for technical support.

**Author contributions:**

K.S.E. E. conceived the concept of using multiple-delay, amplified frequency comb pulse pairs for precision spectroscopy. J. M. developed the laser system, performed the measurements and the data analysis; I.B. set up the spectroscopy part of the experiment. J. M. developed the theoretical framework of Ramsey-comb spectroscopy and wrote the paper together with K.S.E. E., who also supervised the project.

**Competing financial interests:**

The authors declare no competing financial interests.

# Supplementary information: Ramsey-comb spectroscopy with intense ultrashort laser pulses

Jonas Morgenweg, Itan Barmes, Kjeld S. E. Eikema*

LaserLaB Amsterdam, VU University, De Boelelaan 1081, 1081HV Amsterdam, The Netherlands.
*Correspondence to: K.S.E.Eikema@vu.nl

**The atomic phase evolution**

The time-domain analysis of Ramsey-comb spectroscopy relies on tracking the phase-evolution of the recorded upper state population signal. It is instructive to construct the multi-transition situation starting from the single resonance case, which simply exhibits a linear phase evolution (Fig. S1, first column). The second column of Fig. S1 depicts the situation when an equally strong second resonance is added. Now the superimposed signal exhibits a beating pattern, which results in phase jumps every time the signal envelope goes to zero. This phenomena is well-known from the field of acoustics, where the superposition of two similar acoustic frequencies (e. g. two slightly detuned tuning forks) produce a modulation of the sound amplitude according to the frequency difference of the involved sound waves. While in the special case of two transitions of equal amplitude the relative phase is still constant (apart from periodic phase jumps), the situation changes when the spectral amplitudes are unequal (Fig. S1, third column). Finally, adding further transitions leads to a complex phase evolution pattern as depicted in the last column of Fig. S1. In this characteristic phase trace, however, the full time domain information of the signal is encoded. Thus measuring the phase evolution of the signal provides sufficient information for the synthesis of the spectral content.

**Fitting the data: Time domain versus frequency domain**

Most spectroscopic methods are based on data analysis in frequency domain, which means that an optical excitation or absorption spectrum is fitted to obtain the transition frequencies of the excited resonances. In general, the spectral domain has the advantage that the individual resonances are, at least to some extent, decoupled if the spectral resolution is high enough. Also in Ramsey-comb spectroscopy the spectrum, as calculated from the time domain Ramsey-signals via a discrete Fourier-transform (DFT), can be used to extract the underlying resonances.

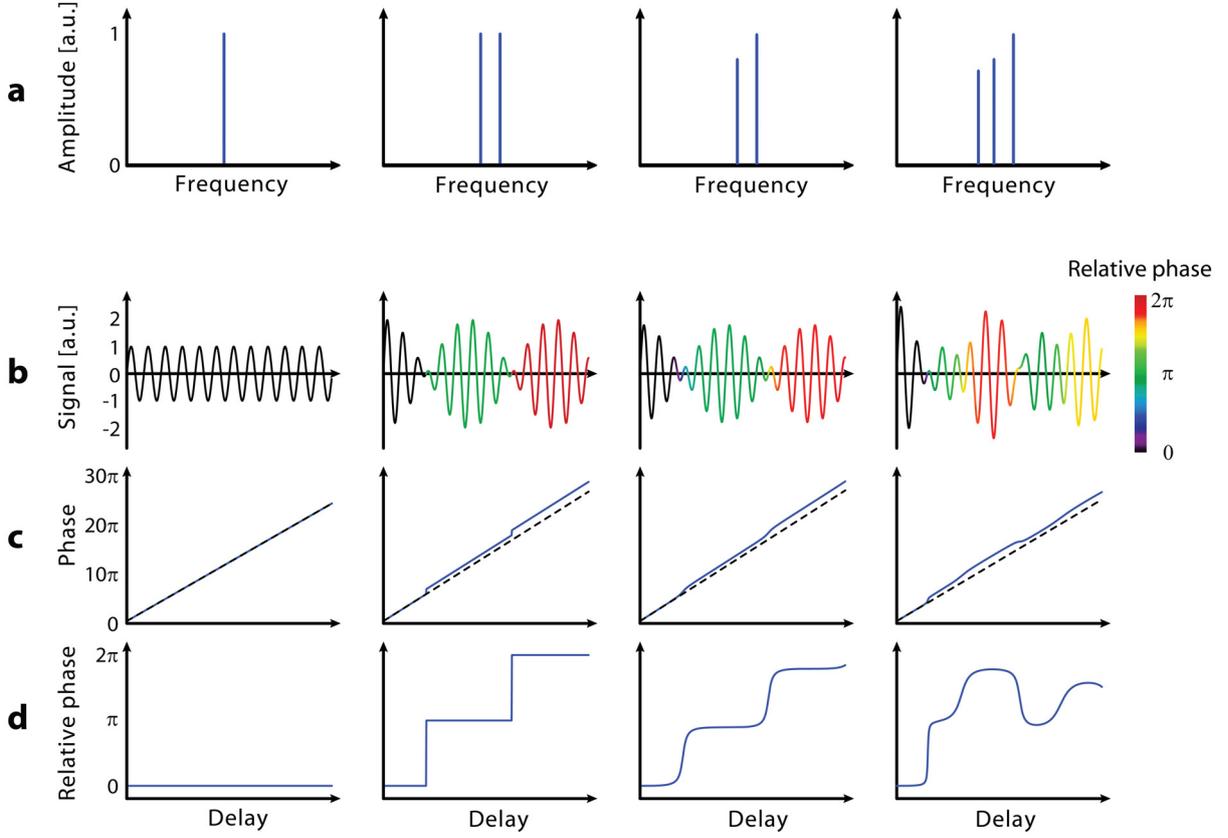

**Figure** S1: **Visualization of the phase evolution of different beating patterns.** Each column describes a different set of parameters. **a,** Spectral amplitudes and frequencies of the atomic transitions. **b,** Time-domain Ramsey-signals for different delays of the excitation pulse pair. **c,** Signal phase. The dashed line represents a linear reference phase. **d,** Phase evolution relative to the reference phase.

However, the analytical description of the spectrum and hence the complexity of the fitting model greatly increases if more than one resonance is excited. This can be understood from looking at the analytical description for the calculated spectrum, which can be derived with the help of basic signal processing theory [1]. For two resonances with transition frequencies $f_1$, $f_2$ and amplitudes $A_1$, $A_2$, the analytical spectrum can be expressed as (for simplification, we neglect here the negative frequency components as well as the influence of the finite scanning length of the individual Ramsey-zones):

$$|DFT(S)|^2 \propto ss_1^2 + ss_2^2 + 2 \cdot \cos[(f_1 - f_2)(N+1)\pi T] ss_1 ss_2 \qquad (2)$$

with

$$ss_i = A_i \frac{\sin[N\pi T(f - f_i)]}{\sin[\pi T(f - f_i)]}, i = 1,2. \qquad (3)$$

The last term on the right side of equation (2) is due to the interference of the two transitions, and depends on the maximum number of Ramsey-zones $N$ and the macro-step delay $T$. This additional interference term makes Ramsey-comb spectroscopy fundamentally different to for example full repetition rate frequency comb spectroscopy, which relies on the superposition of excitation amplitudes instead of upper state populations (proportional to the excitation amplitude squared) as for Ramsey-comb spectroscopy. In practice, spectral line shaping mechanism like for example a finite laser-linewidth, lifetime- and Doppler-broadening further increase the complexity of these spectral interferences. A more elaborate discussion of the analytical description will be published in a successive article.

However, since both the transition frequencies and amplitudes are fully encoded in the phase evolution of the temporal signal, the fitting can be performed purely in the time domain without converting to the frequency domain. The fitting procedure is illustrated in Fig. S2. First an arbitrary frequency $f_0$ is chosen as a reference, e.g. close to the expected position of the measured resonances. The phase of each individual Ramsey-zone is determined relative to the linear reference phase $2\pi f_0 \Delta t$ by sinusoidal fits of the experimentally obtained signals (Fig. S2a).

These relative phases as a function of macro-delay steps are then fitted based on the phase of the analytical time domain signal

$$\Phi_{fit}(\Delta t; A_1, A_2, \ldots; f_1, f_2, \ldots; \Delta\phi) = \arg\left\{\sum_i A_i \exp(-i 2\pi f_i \Delta t)\right\} + \Delta\phi - 2\pi f_0 \Delta t, \qquad (4)$$

including a potential, constant phase shift $\Delta\phi$ as an additional parameter. In the experiment, $\Delta\phi$ incorporates constant phase shifts that might occur during the parametric amplification (which can be up to a few hundred mrad depending on the amplifier alignment) of the frequency comb pulses. However, in Ramsey-comb spectroscopy also a common light-shift (due to the AC-stark effect) simply adds to $\Delta\phi$ and is therefore decoupled from the determination of the transition frequencies. This is a crucial feature, since typically the light shift has to be quantised by repeating the measurement at different power levels and extrapolation to zero excitation power.

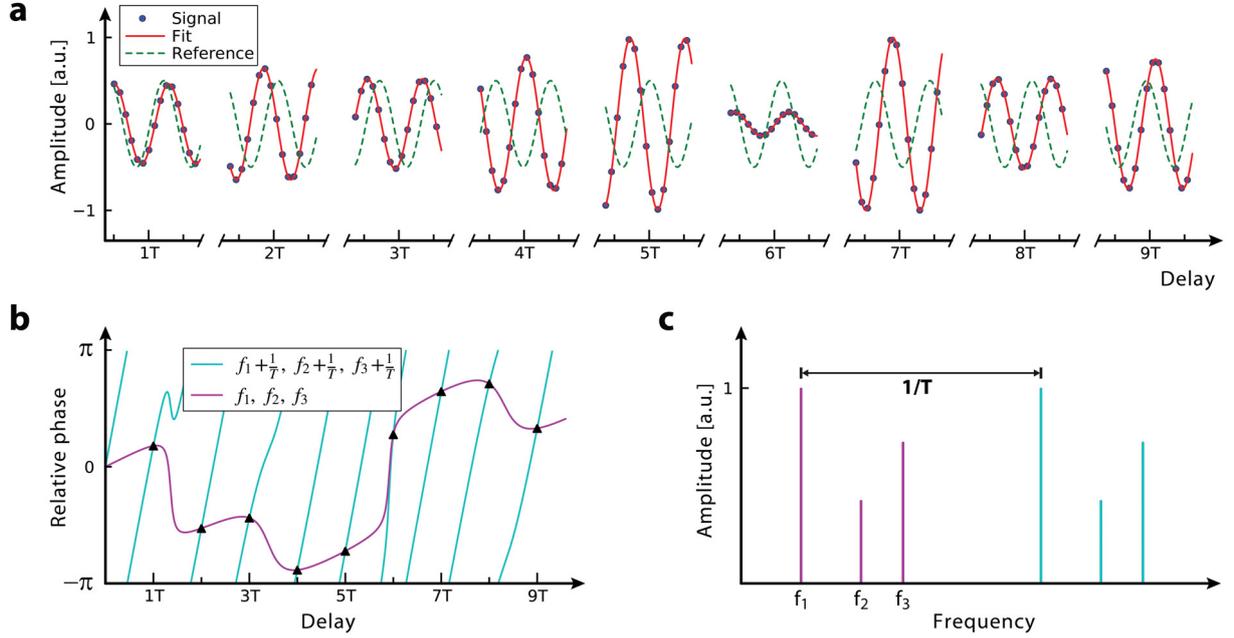

**Figure S2: Visualization of the temporal phase-fitting procedure. a,** Simulated signal (blue dots) together with a sinusoidal fit (red solid line), and the fixed frequency reference trace (green dashed line); $T$ is the macro-delay step between the Ramsey-zones. **b,** The relative phase between the signal and the reference trace. Two possible fits for frequencies off by $1/T$ are shown. **c,** Frequency domain representation of the two possible outcomes from the phase fit.

Because the phase trace is not fully recorded, but only sampled at certain inter-pulse delay steps spaced by $T$, the frequency can only be determined modulo $1/T$. As an example, Fig. S2b shows two groups of frequencies that fit the data equally well, but differ by $1/T$ as depicted in Fig. S2c. In general, this inherent ambiguity can always be solved by repeating the measurement at slightly different delay steps if no previous measurements are available with sufficient accuracy.

**Discussion of the spectroscopic results**

In the presented experiment, three different atomic systems were investigated. Fig. S3 schematically depicts their relevant energy levels. All examined transitions are electric dipole-forbidden between two S-states ($\Delta L = \Delta F = \Delta M_F = 0$), and are excited non-resonantly via a two-photon excitation. The upper and lower energy levels are split up due to hyperfine interaction, and the transitions occur between hyperfine levels of the same $F$-number.

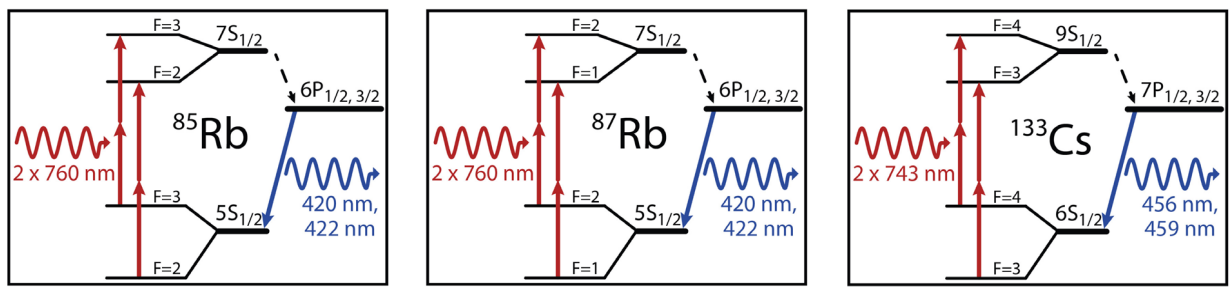

**Figure S3: Level diagrams of the investigated transitions**. The schematics show the relevant levels for the two rubidium isotopes ($^{85}$Rb and $^{87}$Rb) and caesium (only one isotope, $^{133}$Cs). Indicated are the excitation paths in red and the fluorescence light used for detection in blue.

The obtained values for the investigated transitions are the result of averaging over 28 measurement sets (typical recording time for one set: 10 - 15 min). Within one measurement set, Ramsey-fringes of at least 25 different macro-delays are recorded and the transition frequencies of rubidium or caesium are determined with the help of the previously described time-domain fitting procedure. The resulting statistical errors of the transition frequencies of each set are combined with a statistical uncertainty of 20 - 35 kHz, which accounts for random statistical phase shifts due to power fluctuations of the amplifier system. Combining all the measured transition frequencies then leads to the results as shown in Tab. S1.

Also quoted in Tab. S1 are the commonly-used transition frequencies before hyperfine interaction ("centre-of-gravity frequency") and the hyperfine splitting (A) constants, which are both calculated from the measured transition frequencies between the different *F*-states [2]. For rubidium the isotope shift is derived as the difference of the centre-of-gravity frequencies of the two isotopes.

A range of systematic effects has been analysed and taken into account. Table S2 summarises the relevant systematic effects, which lead to the corrected results and systematic uncertainties that are shown in Tab. S1. In the following we briefly discuss the treatment of the individual systematic effects.

| Frequency description | | Final result [kHz] |
|---|---|---|
| $^{85}$Rb, $5S_{1/2}$ - $7S_{1/2}$ | F=2 - F=2 | 788,798,565,751(6)$_{stat}$(4)$_{sys}$ |
| | F=3 - F=3 | 788,795,814,071(5)$_{stat}$(4)$_{sys}$ |
| | Centre of gravity | 788,796,960,604(4)$_{stat}$(4)$_{sys}$ |
| | Hyperfine $A_{7S}$ | 94,684(2)$_{stat}$(0)$_{sys}$ |
| $^{87}$Rb, $5S_{1/2}$ - $7S_{1/2}$ | F=1 - F=1 | 788,800,964,103(9)$_{stat}$(4)$_{sys}$ |
| | F=2 - F=2 | 788,794,768,943(7)$_{stat}$(4)$_{sys}$ |
| | Centre of gravity | 788,797,092,128(6)$_{stat}$(4)$_{sys}$ |
| | Hyperfine $A_{7S}$ | 319,761(6)$_{stat}$(0)$_{sys}$ |
| | Isotope shift $^{87}$Rb - $^{85}$Rb | 131,524(7)$_{stat}$(3)$_{sys}$ |
| $^{133}$Cs, $6S_{1/2}$ - $9S_{1/2}$ | F=3 - F=3 | 806,766,286,786(8)$_{stat}$(4)$_{sys}$ |
| | F=4 - F=4 | 806,757,534,152(7)$_{stat}$(4)$_{sys}$ |
| | Centre of gravity | 806,761,363,429(5)$_{stat}$(4)$_{sys}$ |
| | Hyperfine $A_{9S}$ | 109,999(3)$_{stat}$(0)$_{sys}$ |

**Table S1: Final spectroscopic results including systematic shift corrections.** The statistical and systematic uncertainties are shown in brackets (standard "1-σ-errors", denoting a 68% confidence interval). Note that most of the systematic errors cancel when calculating the Hyperfine A constants, resulting in small systematic errors <0.5 kHz.

| | Blackbody radiation | 2$^{nd}$ order Doppler shift | 2$^{nd}$ order Zeeman shift | Pressure shift | Amplifier phase shift | Residual AC-stark | Total |
|---|---|---|---|---|---|---|---|
| **Rb** | -0.6(0.0) | -0.4(0.0) | -3.5(0.8), $^{85}$Rb F2-2<br>2.5(0.6), $^{85}$Rb F3-3<br>-1.7(0.4), $^{87}$Rb F1-1<br>1.0(0.2), $^{87}$Rb F2-2 | 1.5(2.7) | 0.0(2.5) | (1.0)* | -2.9(3.9), $^{85}$Rb F2-2<br>3.1(3.9), $^{85}$Rb F3-3<br>-1.0(3.8), $^{87}$Rb F1-1<br>1.6(3.8), $^{87}$Rb F2-2 |
| **Cs** | -0.4(0.0) | -0.2(0.0) | -2.3(0.5), $^{133}$Cs F3-3<br>1.8(0.4), $^{133}$Cs F4-4 | -0.6(3.0) | 0.0(2.5) | (2.0)* | -3.5(4.4), $^{133}$Cs F3-3<br>0.5(4.4), $^{133}$Cs F4-4 |

**Table S2: Overview of systematic shifts and uncertainties.** All quoted values are in kHz. *For the AC-Stark shift effect only the uncertainty is given; each individual measurement set was corrected separately for the residual AC-stark shift (see explanation in text).

*Blackbody radiation.* Dynamic Stark shifts due to blackbody radiation are estimated from extrapolating the calculations by Farley and Wing [3] to the measured temperature of the atomic vapour of 50(5)°C.

*Doppler shift.* The counter-propagating excitation scheme suppresses the first order Doppler shift. The second order Doppler shift of the transition frequency $f_{tr}$ is calculated as $\delta f = f_{tr} v^2 / 2c^2$ [4], where $c$ the speed of light, and $v$ is the average speed of the atoms (284(2) m/s for $^{85}$Rb, 281(2) m/s for $^{87}$Rb, and 227(2) m/s for $^{133}$Cs), based on the Boltzmann-distribution for rubidium and caesium vapour at a temperature of 50(5)°C.

*Magnetic (Zeeman) shift.* The investigated S-S transitions are inherently insensitive to first order for magnetic shifts. The second-order magnetic shift depends on the hyperfine quantum number $F$ and can be calculated using second-order perturbation theory [5, 6]. The shifts shown in Tab. S5 are based on a measured magnetic field of 0.85(0.10) G in the interaction zone of the gas cell.

*Pressure shifts.* In order to estimate potential frequency errors that depend on the vapour pressure of the atomic gases, the cell was temporarily heated to >110°C, resulting in more than 100 times higher vapour pressures than typically used during the experiment. Based on frequency measurements at these high pressures (with reduced accuracy due to detrimental effects from the high vapour pressure), the potential frequency error was linearly extrapolated to the typical vapour pressure during the experiments (according to a vapour temperature of 50(5)°C).

*Amplifier phase shift error.* In a separated measurement, we measured the potential delay-dependency of the amplifier phase shift using spectral interferometry analogous to [7]. With the help of an improved measurement over more than 600 ns of inter-pulse delay (in [7] only pulse delays of ~330 ns were investigated), the amplifier phase shift was found to be constant within the accuracy of the phase measurement of 5 mrad (<1/1000th of an optical cycle). Assuming a linear relation between phase shift and delay, this leads to a conservative uncertainty of 2.5 kHz for potential frequency errors due to systematic amplifier phase shifts.

*Residual AC-stark shift.* Ramsey-comb spectroscopy is inherently insensitive to constant phase shifts, which also includes the AC-stark shift caused by the interaction pulses. However, if during one measurement set the absolute pulse energy of the excitation pulses systematically drifts, this can cause a small residual phase change and hence frequency errors. Therefore we a priori measured the phase shift as a function of excitation pulse energy for both rubidium and caesium. Together with the pulse energy drift, which was obtained from a linear fit of the recorded pulse energies during the measurement, the individual fit results could then corrected. The applied corrections were typically <3 kHz for rubidium and <6 kHz for caesium. We estimate the uncertainty of this residual AC-stark shift correction to be 1.0 kHz in the case of rubidium and 2.0 kHz for caesium.